APFA 6 - Applications of Physics in Financial Analysis 6th International Conference 4 - 7 July 2007, Lisbon, Portugal

# BIDDING STRATEGY WITH FORECAST TECHNOLOGY BASED ON SUPPORT VECTOR MACHINE IN ELECTRCITY MARKET

C. Gao[1], E. Bompard[2], R. Napoli[2], Q. Wan[1]
1. Southeast university –School of Electrical Engineering, 210096, Nanjing, Jiangsu, P.R.China.
2. Politecnico di Torino–Dipartimento di Ingegneria Elettric 10129, Torino, Italy

**Abstract:** The participants of the electricity market concern very much the market price evolution. Various technologies have been developed for price forecast. SVM (Support Vector Machine) has shown its good performance in market price forecast. Two approaches for forming the market bidding strategies based on SVM are proposed. One is based on the price forecast accuracy, with which the being rejected risk is defined. The other takes into account the impact of the producer's own bid. The risks associated with the bidding are controlled by the parameters setting. The proposed approaches have been tested on a numerical example.

## I Introduction

The electricity market is widely adopted in the world since its debut in Chile, 1982. With the electricity market, the price of the electricity is no more defined by the monopolist, which operates the power industry before the deregulation. The monopoly system breaks down into an oligopoly market. The players bid in the market following various strategies to maximize their utilities and therefore determine the market behaviors. Many literatures have studied the bidding strategies and their impacts to the market behaviors.

Game theory is a useful tool to study the interactions between the players. The major feature of the game theory is that it takes into account the effect that the decision of one player will influence the decision of the others. It fits the context of the electricity market, hence has been widely applied for analyzing the behaviors of the market players [1][2].

But game theory application needs a lot of detailed information about the market, while it is rather difficult to capture the exact situation of the market and other participants; especially a small changes even in the mind of one participant will possibly result in huge difference in market result, hence game theory is good for theoretically explaining an electricity market, providing a reasonable market behavior, but not for predicting the exact market outcome. The practical bidding strategy should be able to control the risk of the bidding; the '*equilibrium*' of the game in the electricity market is usually an isolated point, which is in some extent very arbitrary in terms of various information scenarios (perfect information is not realistic) and willingness of the market participants (e.g. maximize the market share or profit). In literature, the study on bidding strategy is concentrated on the theoretical analysis of the market behavior, but seldom on making the practical bidding strategy.

The forecasted price is claimed to provide the references for the market bidding and has been intensively investigated with various technologies [3][4][5]. But there lacks of the study on how to make use of the forecast. In [3], the bidding strategy is simply to offer a price a little cheaper than the MCP (Market Clearing Price). It seems too rough and as it is indicated in [3], to apply this strategy, the player should be a market follower, namely what the player bid should not influence the market price. It is obviously not true in an oligopoly market. When the market players are making their bids, the impacts of their own biddings on the market price cannot be omitted.

No matter with what kind of the bidding strategy, it is the major concern of the players to collect the useful market information, including the load information, the history prices and the information about their competitors' generation costs and unit commitments, based on which, the optimal bidding strategies are defined.

Support Vector Machine (SVM) is an excellent machine learning tool that is able to sensitively capture the evolution of the market price[6]. In this paper, we developed two bidding strategies, which are firmly grounded on a machine learning based price forecast or surplus forecast technology. With this strategy, the bidder is able to evaluate his currently status with respect to the market share concentration and chooses its optimal bidding prices considering the associated risks. Monte-Carlo method is used to simulate the market context, which refers to the load evolution and the bidding of the other market participants.

The remainder of the paper is composed by the following sections. In section II, the market clearing mechanism in the electricity market is briefly introduced; in section III, the price forecast based on SVM is explained and the bidding strategies based on price forecast and surplus forecast are presented in section IV; the proposed bidding strategy is tested by the numerical examples in section V and finally conclusions are drawn in section VI.

## II Electricity market modeling

Before deregulation, the power system composed by generation, transmission and distribution systems is usually operated by a monopolist, which is strictly regulated by the authority. The adoption of the electricity market is to foster the competition so as to improve the efficiency. This competition is usually introduced to the power supply and demand side. Due to the technique reason, the transmission sector remains the system of monopoly and is required to be non discrimination, open access to accommodate a fair compete field to the market participants. There will be an independent system operator who manages the system operation following the market outcome with some necessary revision due to the technique constraints. The players of the electricity market can make their transactions through bilateral negotiation or bidding in the power exchange (PX). Therefore the price can be privately determined or defined by the bidding of the players. The former is not publicly declared; the latter is known by every one and will be a reference for the former. In literature the price forecast is for PX price and the study on bidding strategy is for PX transaction as well, although in reality most electricity is transacted by BT.

The bids of the players can be a function representing the relation between the price and quantity (1 in Fig.1) or a pair of quantity and price (2 in Fig.1).

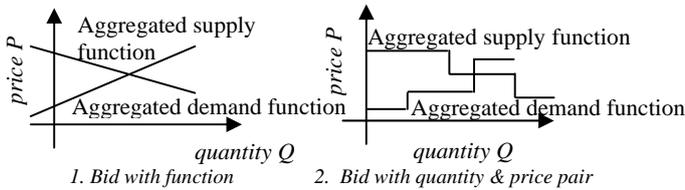

Fig. 1 market clearing in the electricity market

## III Price Forecast Technology with SVM

The electricity price is the bidding result of the market participants. Since the players arbitrarily make their bid, some scholars even claim that the electricity price cannot be predicted [7]. The market players aim to maximize their utility, therefore game theory is suitable to predict the behavior of the market participants as well as the market results. But it is still hard to capture the players' thought in reality, because the information is usually incomplete and sometimes the dominant player won't fully excise the market power to avoid the strict regulation. Time series and statistic methods are also very popular for predicting the market price, the major drawback is that the impact of the specific player's behavior on the market price is hard to be incorporated into the time series model. Machine learning is a subfield of the artificial intelligence and is capable of dealing with the complexity problem. It includes the supervised learning and the unsupervised learning, the former has a definite correct result to follow and the latter hasn't but to pursue a better result. Multi-agent system is a kind of unsupervised learning theory, which is already used for making the bidding strategy [8]. SVM and ANN are the supervised learning approaches, which have been used for predicting the market price, but lack of the research on making the bidding strategy. In this paper, SVM is proposed for determining the bid strategy.

Generally speaking, SVM is to minimize the structural risk instead of the usual empirical risk by minimizing an upper bound of the generalization error and obtains excellent generalization performance [9][12]. Moreover, SVM has already been used for classification [13][14], regression[15][16] and time series prediction [17][18]. SVM is well known for the small sampling problem and has shown excellent performance in predicting the electricity market price [6].

SVM is to map the input data $x$ into a higher dimensional feature space through a nonlinear mapping $\Phi$ and then a linear regression problem is obtained and solved in this feature space. With the given training data $\{(x_1,y_1), \ldots (x_i,y_i), \ldots (x_n,y_n)\}$, the mapping function can be formulated as,

$$f(x) = \sum_{i=1}^{n} \omega_i \Phi_i(x_i) + b \quad (1)$$

where $\omega_i$ and $b$ are the parameters need to be defined. $\varepsilon$-SVR is to find a function $f(x)$ that has at most $\varepsilon$ deviation from the actually obtained targets $y_i$ for all the training data and at the same time is as flat as possible. Flatness in this case means to reduce the model complexity by minimizing $\|\omega\|^2$, we can write this problem as an optimization problem:

$$\text{Min } \frac{1}{2}\|\omega\|^2$$
$$s.t. \begin{cases} y_i - \Phi(\omega, x_i) - b \leq \varepsilon \\ \Phi(\omega, x_i) + b_i - y \leq \varepsilon \end{cases} \quad (2)$$

which means we do not care about errors as long as they are less than $\varepsilon$, but will not accept any deviation larger than this. To be more realistic, one can add slack variables $\xi_i$, $\xi_i^*$ $i=1,2,3\ldots n$, to cope with otherwise infeasible constraints of the optimization problem (2). Hence we arrive at the formulation stated in [11]:

$$\text{Min } \frac{1}{2}\|\omega\|^2 + C\sum_{i=1}^{n}(\xi_i + \xi_i^*) \quad (3)$$

$$(3\text{-}1)$$

$$(3\text{-}2)$$

$$s.t. \begin{cases} y_i - \Phi(\omega, x_i) - b \leq \varepsilon + \xi_i \\ \Phi(\omega, x_i) + b_i - y \leq \varepsilon + \xi_i^* \\ \xi_i, \xi_i^* \geq 0 \end{cases}$$

where $C$ is a positive constant (known as regularization parameter). The optimization formulation can be then transformed into a dual problem (Vapnik, 1998)[12] and the solution is expressed as

$$f(\mathbf{x}) = \sum_{i=1}^{n}(\alpha_i - \alpha_i^*)K(x_i, x) + b \qquad (4)$$

where $\alpha_i, \alpha_i^*$ are the dual variables with reference to constraints (3-1) and (3-2) and $0 \leq \alpha_i, \alpha_i^* \leq C$. the constant $C$ is the trade off between the flatness of $f$ and the amount up to which deviation larger than $\varepsilon$ are tolerated.

$K(x_i, x_j) = \Phi(x_i) \cdot \Phi(x_j)$ is the kernel function that performs the non-linear mapping, which must satisfy the Mercer's conditions. Those sample points that appear with non-zero coefficients in (4) are so called support vectors (SV). Kernel function can be Gaussian function, polynomial function, sigmoid function etc.; In our SVR model, we apply the Gaussian kernel as (5), the convenience of which has been demonstrated in[19][20];

$$K(x_i, x_j) = \exp(-\frac{\|x_i - x_j\|}{2p^2}) \qquad (5)$$

SVM makes the mapping between a series of input and the market price. The input should be able to characterize and account for the variation of the price. For instance, in Fig.2 the load value $L_t$, the day type $T$ and the hour $t$ are taken as the factors to drive the price. By SVM, the mapping function between these factors and the market price are identified. From the outsider point of view, it would be appropriate to forecast the market price. But for the market participants, their own bids should be considered as an important factor, since their bidding strategies are able to cause the price variation.

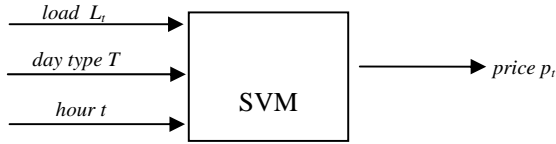

*Fig.2 Price forecast by SVM*

## IV Bidding Strategies Based on Price Forecast and Surplus Forecast

In electricity market, the producer's minimum bidding price is its marginal cost. It is also its optimal choice under perfect competition, in which the market players are price-takers. Sometimes the offering quantity of a generator is divided into two parts; the first one is corresponding to the minimum power output, which cannot be switched off due to the technique reason. To ensure its being dispatched, zero price is offered. The second offer will be considered for strategic gaming. Since the first offer can be directly subtracted from the original load as the bilateral contract and for sake of simplicity, we will not consider the technique constraints such as the ramp rate limit of the switch on and off.

We propose two SVM based models for the market players making their bidding strategies. In the first approach, the bid is defined considering the rejection risk with reference to the forecasted price. In the second approach, the player's surplus function is formulated by SVM, its optimal bid is obtained by surplus maximization.

*A. Bidding with price forecast*

The market price at hour $t$ can be represented by:
$$p_t = p_t^F + \varepsilon_t \qquad (6)$$
$$\varepsilon_t \sim N(\mu_t, \sigma_t^2) \qquad (7)$$

where, $p_t$ is the actual price at the time $t$, $p_t^F$ is the price output of the SVM forecast for point $t$, $\varepsilon_t$ is a stochastic variable subject to the normal distribution with mean $\mu_t$ and variance $\sigma_t^2$, which are obtained from the historical forecast.

Under an oligopoly market, the players are willing to push prices up; hence have the motivation to offer a high price with an associated acceptable risk.

As reported in Fig.3, we suppose the producer will bid the price $p_b$ and the related acceptance probability is $p(p_t > p_b)$. The bid is determined with a certain acceptable confidence $p(p_t > p_b) = \alpha$, which depends on the player's attitude to the risk. Therefore the optimal bidding price can be formulated as:

$$p_b^{opt} = \begin{cases} p_b^{\alpha} & p_b^{\alpha} \geq C^m, p(p_t \geq p_b^{\alpha}) = \alpha \\ C^m & p_b^{\alpha} \leq C^m \end{cases} \qquad (8)$$

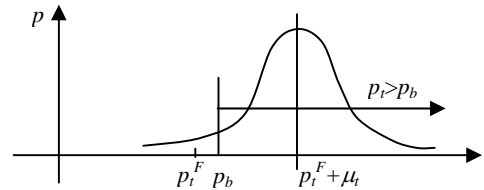

*Fig.3. Market price distribution with price forecast result expression*

The model helps the market players make their biddings and provides them a way to control their risks. For the player with small market share, usually they are more conservative, they may just offer the marginal cost to have the maximum possibility to be accepted, accordingly, $\alpha$ can be relatively high. For the bigger players, they are more powerful in determining the market price hence are more likely to accept higher risk, although with this approach, the incentive of accepting higher risk remains unclear.

## B. Bidding with surplus forecast

The goal of the market player defining their bidding strategies is to maximize their producer surplus, which can be formulated as:

$$S = p_t * q - C(q) \quad (9)$$

where $q$ is the accepted quantity, $C$ is the cost function of the producer.

The offer of the producer may impact the market price. This impact is not considered in the previous model and in some cases that will significantly influence the accuracy of the forecast as well as the definition of the bidding strategy. Therefore, the player's own bid should be taken into account as an important input, which is also the basis for making the bid. As shown in Fig.4, considering all the driving factors, the producer surplus can be expressed as:

$$S = f(L_t, T, t, p_b) \quad (10)$$

(10) implies that the complicated formulation of the market price $p_t$ and the dispatched quantity $q$ are implicitly expressed in the surplus function $f(L_t, T, t, P_b)$, which can be identified by SVM.

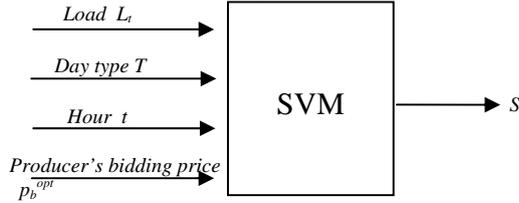

*Fig.4. Surplus forecast for players with significant share*

On the other hand, from theoretical point of view, considering more meaningful inputs that can account for the variation of the output, we can forecast the price better. For the specific next hour $t$, since that the short term load forecast is with high accuracy, we assume that the load $L_t$ is known, hence the surplus function has only one decision variable $p_b$, which need to be defined by the player. The producer aims to maximize his surplus, which is expressed as:

$$\text{Max } S = f(p_b) \quad (11)$$
$$s.t. \; p_b^{min} \leq p_b \leq p_b^{max} \quad (12)$$

where $p_b^{max}$ and $p_b^{min}$ are respectively the maximum and minimum bidding price in the sampling data; the sampling data are required to be timely valid, namely the latest data are with more weights. Suppose $p_0$ is the bidding price corresponding to the maximum surplus with reference to (11). The optimal bidding price $p_b^{opt}$ can be defined as Tab.1.

When $p_b^{min} < p_0 < p_b^{max}$, SVM can map the input($L_t$,T,t, $p_b$) and output($S$) well, the producer can bid $p_b^{opt}$ exactly equal to $p_0$.

When $p_0 = p_b^{min}$, and sometimes the surplus is zero, it means that the player's bid is possibly the marginal offer. To bid lower price may result in more quantity accepted.

Therefore, $p_b^{opt}$ keeps $p_b^{min}$ with probability $\gamma_1$% and there will be a $\beta_1$%$p_b^{max}$ decrease with probability 1-$\gamma_1$%. Higher $\beta_2$% the player is more conservative in influencing the market price.

when $p_0 = p_b^{max}$, the player has the motivation to tentatively bid a higher price (1+$\beta_2$%) $p_b^{max}$, therefore $p_b^{opt}$ keeps $p_b^{max}$ with probability $\gamma_2$% and there will be a $\beta_2$%$p_b^{max}$ mark up with probability 1-$\gamma_2$%. $\beta_2$% is a parameter representing the player's attitude to the risk. Higher $\beta_2$% implies more confident in determining the market price.

*Tab.1 Definition of $p_b^{opt}$ for producer*

| $p_0$ | $p_b^{opt}$ | probability |
|---|---|---|
| $p_b^{min}$ | $p_b^{min}$ | $\gamma_1$% |
| | $p_b^{min}(1-\beta_1\%)$ | 1-$\gamma_1$% |
| $p_b^{min} < p_0 < p_b^{max}$ | $p_0$ | 1 |
| $p_b^{max}$ | $p_b^{max}$ | $\gamma_2$% |
| | $p_b^{max}(1+\beta_2\%)$ | 1-$\gamma_2$% |

## V Numeric examples

The proposed two approaches are applied to a test example, in which the load follows the profile of the New England market in 2004(Fig.5). It would be difficult to invent a specific real market context, because the market players may adopt various bidding strategies, and the attitude toward the risk and the incomplete information will contribute a lot to the variation of the bidding strategies. We would like to create a context, in which the players bid with reasonable strategies. The capacity, the marginal cost and the original parameter $\alpha$ are detailed in Tab.2. Highlighting the two approaches and better find the impact of the bidding strategy, we define that the producers have the same marginal costs. We will study the market evolution in 120 days, the first 30 days will be the preliminary stage and in the next 90 days the producers' strategic bidding will be examined.

Because the electricity market is an oligopoly market, the market concentration is usually high. The HHI index of the market is designed to be 1702.2, which means that it is close to a highly concentrated market[*]. The parameter $\alpha$ of each player in the base case is defined with reference to his market share, lower market share, the player is less confident in making the market price, more conservative in bidding, and there will be a higher $\alpha$ corresponded.

---

[*] The HHI of a market is calculated by summing the squares of the percentage market shares held by the respective firms. For example, an industry consisting of two firms with market shares of 70% and 30% has an HHI of 70²+30², or 5800. People regard a market, in which HHI is below 1000 as "unconcentrated," between 1000 and 1800 as "moderately concentrated," and above 1800 as "highly concentrated."

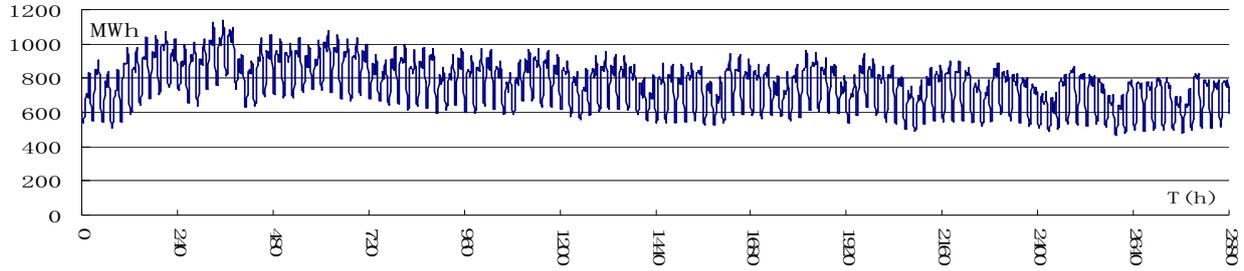

*Fig. 5 Load profile of the test case, which follows the profile of New England market in 2004*

*Tab.2 The data of the companies in the market*

|  | C1 | C2 | C3 | C4 | C5 | C6 | C7 | C8 | C9 | C10 | C11 | C12 | C13 | C14 | C15 | C16 |
|---|---|---|---|---|---|---|---|---|---|---|---|---|---|---|---|---|
| MC(€/MWh) | 30 | 30 | 30 | 30 | 30 | 30 | 30 | 30 | 30 | 30 | 30 | 30 | 30 | 30 | 30 | 30 |
| Quantiy(MWh) | 15 | 15 | 15 | 50 | 45 | 380 | 380 | 60 | 150 | 60 | 51 | 39 | 28 | 32 | 60 | 20 |
| $\alpha$ | 0.95 | 0.95 | 0.95 | 0.9 | 0.95 | 0.8 | 0.8 | 0.95 | 0.9 | 0.95 | 0.9 | 0.95 | 0.95 | 0.98 | 0.85 | 0.95 |

*Tab.3 Market simulation under various scenarios*

|  | C1 | C2 | C3 | C4 | C5 | C6 | C7 | C8 | C9 | C10 | C11 | C12 | C13 | C14 | C15 | C16 | Σ |
|---|---|---|---|---|---|---|---|---|---|---|---|---|---|---|---|---|---|
| $S_1$(M€) | 0.67 | 0.70 | 0.69 | 2.26 | 2.10 | 16.2 | 16.3 | 2.79 | 6.78 | 2.82 | 2.36 | 1.81 | 1.35 | 1.44 | 2.68 | 0.95 | 61.8 |
| $S_2$(M€) | 1.73 | 1.69 | 1.72 | 5.65 | 4.92 | 37.2 | 35.6 | 6.41 | 16.0 | 6.66 | 5.88 | 4.34 | 3.21 | 3.52 | 6.87 | 2.25 | 143.7 |
| $S_3$(M€) | 1.71 | 2.50 | 1.66 | 5.52 | 4.97 | 34.4 | 37.0 | 6.57 | 16.2 | 6.45 | 5.49 | 4.37 | 3.18 | 3.56 | 6.51 | 2.25 | 142.5 |
| $S_4$(M€) | 1.34 | 2.44 | 1.32 | 4.43 | 3.97 | 46.1 | 27.0 | 5.34 | 12.3 | 5.36 | 4.49 | 3.49 | 2.49 | 2.87 | 5.29 | 1.84 | 130.1 |
| $S_5$(M€) | 0.57 | 1.38 | 0.55 | 1.75 | 1.66 | 14.6 | 34.1 | 2.31 | 5.27 | 2.32 | 1.88 | 1.53 | 1.06 | 1.19 | 2.22 | 0.71 | 73.06 |
| $S_6$(M€) | 0.13 | 0.13 | 0.13 | 0.43 | 0.39 | 1.02 | 1.02 | 0.52 | 1.01 | 0.52 | 0.44 | 0.34 | 0.24 | 0.28 | 0.52 | 0.17 | 7.30 |
| $S_7$(M€) | 1.32 | 1.32 | 1.32 | 3.83 | 3.93 | 11.2 | 8.90 | 5.20 | 10.1 | 5.20 | 3.90 | 3.41 | 2.46 | 2.81 | 3.39 | 1.76 | 69.98 |
| $S_8$(M€) | 1.62 | 1.62 | 1.62 | 4.21 | 4.73 | 16.5 | 15.1 | 6.23 | 10.2 | 6.23 | 4.28 | 4.13 | 2.99 | 3.53 | 3.12 | 2.15 | 88.16 |
| $S_9$(M€) | 1.28 | 1.34 | 1.34 | 3.92 | 3.97 | 10.0 | 8.73 | 5.25 | 10.5 | 5.25 | 3.99 | 3.45 | 2.49 | 2.87 | 3.42 | 1.79 | 69.6 |
| $S_{10}$(M€) | 1.68 | 1.68 | 1.68 | 4.07 | 4.73 | 18.8 | 17.2 | 6.12 | 7.84 | 6.12 | 4.14 | 4.16 | 3.05 | 3.67 | 3.88 | 2.21 | 91.05 |

## A. Preliminary stage

Since we want to make use of the forecast, we should have the historical data. It is exactly what happened in the real market, at beginning, the biddings are tentatively, the players are trying to familiar with the market as well as their competitors. In this period, we assume that the players bid randomly within an interval, which could be between the marginal production cost and the price cap, which is 200€/MWh in the extreme case. The surpluses of each producer may vary drastically. Monte-Carlo method will be applied to simulate the bidding of the producers.

Fig.6 presents market price evolution of the first 5 days, the market price is very irregular.

In Tab.3, $S_1$ is the producer surplus at the preliminary stage (the first 30 days). $S_2$ is the producer surplus in the next 90 days, but the players still bid randomly, the market result is taken as the reference case for the following case (sub-section *B*). The producer surplus are basically defined by the quantity, the producer with more quantity offered will have more surplus. In this period, the producers are getting familiar with the market, preparing for the strategic bidding.

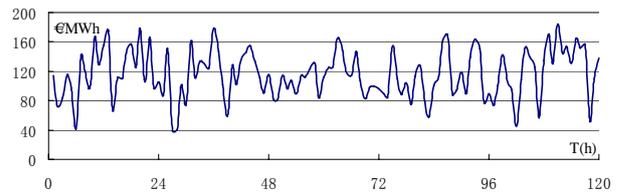

*Fig.6 Market price evolution with randomly bidding at the preliminary stage*

## B. Bidding with price forecast

The producers may bid based on the price forecast, they can control their risks by adjusting the parameter $\alpha$. The impact of the offering quantity can be assessed as well.

1) A small producer bids based on price forecast

After the first 30 days, suppose, there is only one producer C2 bid based on the price forecast with $\alpha$=0.95. The surplus is reported in Tab.3 as $S_3$. Obviously, it is able to gain much more surplus than the C1(1.71M€) and C2(1.66M€). But the market price is almost not influenced, and keeps a very irregular profile as well. Fig. 7 shows the market price between 115th and 120th day. The surplus is reported in Tab.3 as $S_3$, the surplus sum of all the producers keeps almost unchanged (142.5) from the reference case (143.7).

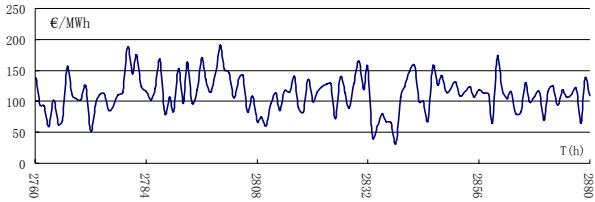

*Fig.7 Market price evolution with C3 strategic bidding based on price forecast*

2) one big producer bids based on price forecast

The producer C6 will bid the price based on the forecasted price with the parameter $\alpha$=0.8. The producer surplus is reported in Tab.3 as $S_4$. C6 acquires much more surplus (46.1M€) than C7 acquires (27.0M€), although they are identical with reference to the marginal cost and offering quantity. But the sum of all the producers' surplus is 130.1 M€, which is decreased compared to the former scenario (142.5M€); it implies that most of the producers suffer from the bidding strategy of C6. If C7 bids based on price forecast with $\alpha$=0.95, the corresponding producer surplus is $S_5$ in Tab.3. C7 gets 134% more surplus than C6, it seems higher $\alpha$=0.95 may result in higher surplus. But the sum of the producer surplus decreases a lot compare to the previous cases, even for C6, it gets less surplus than that in the case the producers bid randomly.

3) all the producers bid based on price forecast

It seems that for a single producer, it will be benefited from the price forecast bidding strategy and higher $\alpha$ will help the producer gain more surpluses. We would like to study how the market behaves under the scenario with all producers taking the price forecast strategy with high parameter $\alpha$ equal to 0.98. In Tab3, $S_6$ reports the surplus result under this scenario. The sum of producer surplus is 7.30M€, which is drastically decreased compared to that of the reference case (143.7M€). Fig. 8 shows the price evolution, we find that after some time, the market price goes down until it reaches 30€/MWh, which is exactly the marginal cost of all the producers. It implies that a perfect competition market is made.

The small producer can bid with the price forecast model; it does not influence too much market result, but can significantly improve its own surplus. But for the big producer, bidding with price forecast tends to result in lowering the market price and makes a perfect competition market.

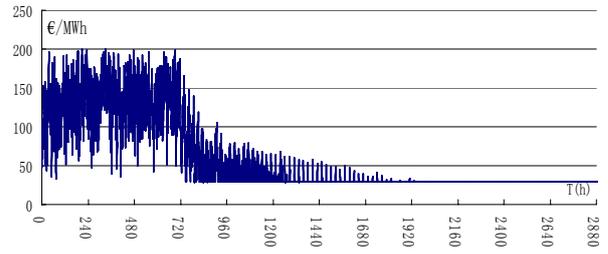

*Fig.8 Market price evolution with all producers strategic bidding based on price forecast with $\alpha$=0.98*

The price forecast bidding strategy does not consider the bidding impact, namely with this strategy, the producer is not able to know how its bidding will impact the market, the major idea is to get its offer accepted by lowering the bidding price while the market price is not concerned. This may be true with the small producers, but in case that the producer is able to significantly influence the market price, it may not adopt this strategy, because its surplus can be improved even with accepted quantity decreased when the market price is increased.

*C. bidding with surplus forecast*

A More effective way is to forecast the surplus directly. SVM tries to map the surplus and the input bidding. It will be a little difficult to solve the optimization problem (11), but as a practical solution, we can compute a certain number of $S$ between $p_b^{min}$ and $p_b^{max}$, inside which the mapping should be accurate. It is reasonable, since the distance between $p_b^{min}$ and $p_b^{max}$ is not so big, for instance 50 or 100, we compute 200 $S$ with $p_b$ uniformly distributed between $p_b^{min}$ and $p_b^{max}$, that should be enough to find the applicable bidding price corresponding to the maximum surplus. The study on the surplus forecast bidding strategy starts from the scenario with all the producers bid with the price forecast bidding strategies, which can be a reference case for the following analysis..

1) one big producer bids based on surplus forecast

In this scenario, only C6 bid with the surplus forecast bidding strategy. Fig.9 shows the price evolution for the last 5 days in the studied time span. The profile is very regular and that is more fitting with very much the price evolution in the real market; we may infer it is made by the big producer C6, who adopt the strategies that are able to keep the market evolve with a certain regular profile with reference to the load profile. Fig. 13 shows that the daily average market price in last 15days marks up a lot from that of the reference case, in which the market price is the marginal cost 30 €/MWh. The strategic bidding of the big producer C6 may account for this price increase. The bids of C6 serve as the marginal units and determine the market prices. $S_7$ in Tab.3 is the related surplus results. All the producers are benefited form the strategic bidding of C6 with reference to $S_6$, with which, the producers follow the price forecast bidding strategy. Compared to the case that all the producers bid according to price forecast, the surplus of each the producer increased a lot.

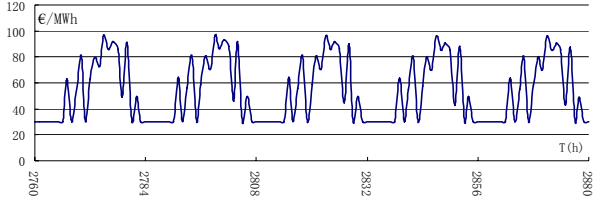

Fig.9 Market price evolution with C6 strategic bidding based on surplus forecast with the parameter $\beta_1=5\%$, $\beta_2=5\%, \gamma_1=80\%, \gamma_2=80\%$ ($115^{th}$-$120^{th}$ day)

2) Two big producer bid based on surplus forecast

Besides the producer C6, C7 also takes the bidding strategy based on surplus forecast. The regular price evolution is shown in Fig. 10. The average daily price of the last 15 days is increased compared to previous scenario as shown in Fig. 13. The result surplus is reported as $S_8$ in Tab.3. The producer surplus of C7 increases from 8.90 M€ to 15.1 M€ with reference to the previous scenario, moreover, the surplus is significantly increased for each producer. It means that as a big producer, it would be good to bid with the surplus forecast approach, which makes its own surplus higher as well as the producer surplus of the other producers.

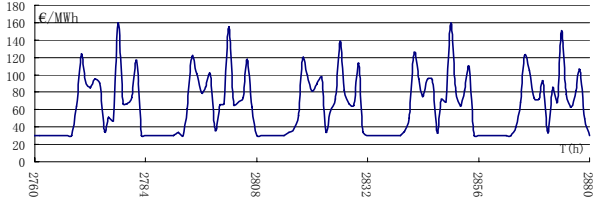

Fig.10 Market price evolution with both C6 and C7 bidding based on surplus forecast with $\beta_1=5\%$, $\beta_2=5\%, \gamma_1=80\%, \gamma_2=80\%$ ($115^{th}$-$120^{th}$ day)

3) One big producer and one small producer bid based on surplus forecast

Besides the producer C6, the small producer C1 also bid with the surplus forecast. The surplus is reported in Tab.3 as $S_9$. The surplus sum of $S_9$ keeps almost unchanged with reference to $S_7$. As shown in Fig. 13, the unchanging of the price is also reflected by the average daily price in the last 15 days with reference to that in the scenario with only C6 bidding with surplus forecast. But the surplus of C1 is less than that of $S_7$, which is the result of only C6 strategically bids based on surplus forecast. It means that C1 cannot improve its surplus by adopting the bidding strategy based on surplus forecast; the surplus of C1 may even worse. Fig.11 presents the market price evolution between $115^{th}$ and $120^{th}$ day under this scenario.

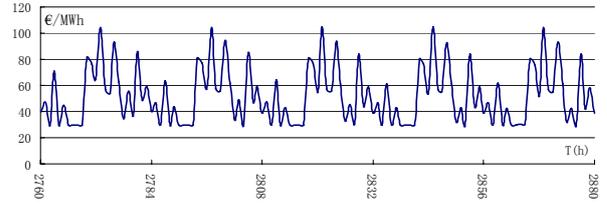

Fig.11 Market price evolution with C1 C6 bidding based on surplus forecast with the parameter $\beta_1=5\%$, $\beta_2=5\%, \gamma_1=80\%, \gamma_2=80\%$ ($115^{th}$-$120^{th}$ day)

4) three producers bid based on surplus forecast

All the three big producers (C6, C7, C9) bid based on maximizing their surpluses. Fig.12 shows the regular price evolution for the last 5 days in the studied time span. The producer surpluses are increased as $S_9$ reported in Tab.3. Moreover, in this scenario, the surpluses of the producers other than the C6, C7, C9 gets almost the same surplus as what they can get in the randomly bidding scenario ($S_2$). The decrease of the surplus sum (from 143.7 M€ to 91.05 M€) is mainly contributed by the loss of three biggest producers.

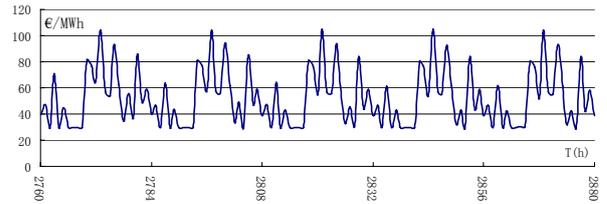

Fig.12 Market price evolution with C6, C7 and C9 bidding based on surplus forecast with $\beta_1=5\%$, $\beta_2=5\%, \gamma_1=80\%, \gamma_2=80\%$ ($115^{th}$-$120^{th}$ day)

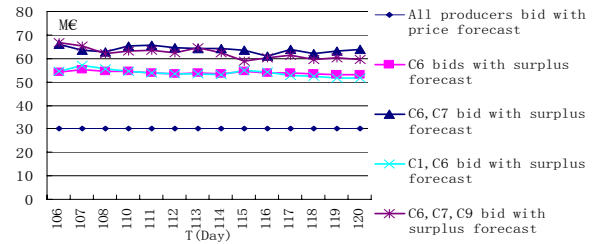

Fig. 13 Daily average price of the last 15 days in various scenarios ($106^{th}$ – $120^{th}$ day)

## VI Conclusion

The electricity market players cannot have complete information, which makes it difficult to define the bidding strategy by accurately computing the market equilibrium. The bidding strategy proposed in this paper is based on forecast technology, which fully makes use of the easily acquired information of the market.

General speaking the price forecast bidding strategy proposed in this paper is a conservative approach; it can be applied by the small producers, which does not aim to increase the market price but to guarantee the acceptance of their offering. In case that the market price is not influenced, this approach can improve their surplus efficiently. It should not be applied by the big producers, since with this approach, their offers may drastically lower the market price and result in surplus loss.

Besides the price forecast bidding strategy for the small producers, we extended the SVM technology to the surplus forecast for the price maker, who owns a significant market share. The players dynamically learn from the market and bid based on the latest acquired information. The variation of the market behavior will be soon perceived and reacted by the proposed bidding strategy. The market share distribution can be effectively reflected by the price evolution. The big producers can get extra surplus by adopting the surplus forecast bidding strategy, but that does not apply to the small producers. The surplus forecast approach takes into account both the effects of the price and quantity acceptance, it is suitable for the big producer to evaluate the comprehensive outcome. But it is hard for a small producer to define the market price; therefore the price mark up mechanism in surplus forecast approach does not help the small producer gain extra surplus but increase its risk of offer rejection.

For a specific producer, it is able to make decisions either by surplus forecast approach or by price forecast approach with various parameters definition, which is able to reflect the producers' attitude to the risk.

Moreover, small producers bid with price forecast and the big producers bid with surplus forecast producer regular price profile, which fits the real market very much. By adjusting the parameters, the proposed approaches are promising for the real electricity market simulation.

## References

[1] Chattopadhyay, D., Multicommodity spatial Cournot model for generator bidding analysis; IEEE Transactions on Power Systems, 19(1), 2004:267 – 275

[2] De la Torre, S.; Contreras, J.; Conejo, A.J., Finding multiperiod Nash equilibria in pool-based electricity markets, IEEE Transactions on Power Systems, 19(1), 2004: 643 – 651.

[3] David, A.K., Fushuan Wen, Strategic bidding in competitive electricity markets: a literature survey, Vol.4, 2002: 2168-2173.

[4] J.P.S. Catalão, S.J.P.S. Mariano, V.M.F. Mendes and L.A.F.M. Ferreira, Short-term electricity prices forecasting in a competitive market: A neural network approach, Electric Power Systems Research, Volume 77, Issue 10, August 2007:1297-1304.

[5] Garcia, R.C., Contreras, J., van Akkeren, M. and Garcia, J.B.C., A GARCH forecasting model to predict day-ahead electricity prices, IEEE Trans. Power Sys., Vol. 20, No.2, May 2005, pp:867 – 874.

[6] Ciwei Gao, Ettore Bompard, Roberto Napoli, Haozhong Cheng, Price forecast in the competitive electricity market by support vector machine, Physica A: Statistical Mechanics and its Applications, 2007, 382(1): 98-113.

[7] F.Schweppe, M.Caramanis, R.Tbors, and R.Bohn, Spot pricing of elelctricity. Norwell, MA: Kluwer, 1998.

[8] Gaofeng Xiong, Tomonori Hashiyama, Shigeru Okuma, An Electricity supplier Bidding strategy through Q-Learning.

[9] V. Vapnik, Chervonenkis, A., "Theory of Pattern Recognition", Nauka, Moscow, 1974.

[10] V. Vapnik, Estimation of Dependences based on empirical data. Springer, Berlin, 1982

[11] V. Vapnik, "The nature of statistical learning theory," Springer-Verlag: New York, 1995.

[12] V. Vapnik, "Statistical learning theory," John Wiley: New York, 1998.

[13] A comparison among four SVM classification methods: LSVM, NLSVM, SSVM and NSVM. Shu-Xia Lu; Xi-Zhao Wang; 2004. Proceedings of 2004 International Conference on Machine Learning and Cybernetics,Vol.7, 26-29 Aug. 2004:4277 – 4282.

[14] J.A.K. Suykens, J.Vandewalle, Least squares support vector machine classifiers. Neural Processing Letters 1999, 9: 293-300.

[15] Alex J.Smola , Bernhard Scholkopf, A tutorial on support vector regression. Statistics and computing 2004, 14:199-222.

[16] Wang, Wenjian; Xu, Zongben, A heuristic training for support vector regression, Neurocomputing Vol.61, Complete, October, 2004, pp. 259-275.

[17] Cao, Lijuan, Support vector machines experts for time series forecasting, Neurocomputing Vol. 51, April, 2003, pp. 321-339

[18] Thissen, U., van Brakel, R., de Weijer, A.P., Melssen, W.J., Buydens, L.M.C., Using support vector machines for time series prediction , Chemometrics and Intelligent Laboratory Systems, Vol.69, Issue: 1-2, Nov. 28, 2003, pp. 35-49.

[19] Keerthi S S, Lin C J. Asymptotic behaviors of support vector machines with Gaussian kernel[J]. Neural Computation, 2003, 15(7): 1667-1689.

[20] Lin H T, Lin C J. A study on sigmoid kernels for SVM and the training of non-PSD kernels by SMO-type methods [R]. Taipei: Department of Computer Science and Information Engineering, National Taiwan University. 2003.